# Limiting optical diodes enabled by the phase transition of vanadium dioxide

Chenghao Wan[1,2], Erik Horak[3], Jonathan King[1], Jad Salman[1], Zhen Zhang[4], You Zhou[5], Patrick Roney[1], Bradley Gundlach[1], Shriram Ramanathan[4], Randall Goldsmith[3], Mikhail A. Kats[1,2]

[1] Department of Electrical and Computer Engineering, University of Wisconsin-Madison, Madison, Wisconsin 53706, USA
[2] Department of Materials Science & Engineering, University of Wisconsin-Madison, Madison, Wisconsin 53706, USA
[3] Department of Chemistry, University of Wisconsin-Madison, Madison, Wisconsin 53706, USA
[4] School of Materials Engineering, Purdue University, West Lafayette, IN 47907, USA
[5] Harvard John A. Paulson School of Engineering and Applied Sciences, Harvard University, Cambridge, MA 02138, USA

**Abstract**

A limiting optical diode is an asymmetric nonlinear device that is bidirectionally transparent at low power, but becomes opaque when illuminated by sufficiently intense light incident from a particular direction. We explore the use of a phase-transition material, vanadium dioxide ($VO_2$), as an active element of limiting optical diodes. The $VO_2$ phase transition can be triggered by optical absorption, resulting in a change in refractive index orders of magnitude larger than what can be achieved with conventional nonlinearities. As a result, a limiting optical diode based on incident-direction-dependent absorption in a $VO_2$ layer can be very thin, and can function at low powers without field enhancement, resulting in broadband operation. We demonstrate a simple thin-film limiting optical diode comprising a transparent substrate, a $VO_2$ film, and a semi-transparent metallic layer. For sufficiently high incident intensity, our proof-of-concept device realizes broadband asymmetric transmission across the near infrared, and is approximately ten times thinner than the free-space wavelength.

**Introduction**

Devices that feature asymmetric transmission of light [1][2][3][4] are useful for protecting delicate photosensitive elements, especially in optical systems with high-power light sources [5][6]. The most common asymmetric optical devices are linear isolators, which utilize the magneto-optic effect to break reciprocity and achieve unidirectional transmission [7][8]. These linear devices are insensitive to the intensity of incident light—which can be either desirable or undesirable depending on the application. Most linear isolators require an external field bias [8][9], and can be difficult to integrate into small-scale optical systems [2].

An alternative approach utilizes optical nonlinearity (*e.g.,* the Kerr effect [10][11]) to achieve intensity-dependent asymmetric transmission. These asymmetric intensity-dependent devices are often referred to as nonlinear isolators or optical diodes (though the latter term is sometimes also used for linear devices). These nonlinear devices can be roughly divided into two types: those that are opaque for low intensities and become transparent when illuminated by sufficiently intense light incident from a particular direction [3][12][13], and those that are transparent at low intensities and become opaque upon high-intensity illumination [4]. In this paper, we focus on the latter type, which for clarity we refer to as *limiting optical diodes*, because they behave as asymmetric optical limiters [14][15]. We note, however, that the same principles and conclusions can be applied to either type of nonlinear isolator.

These nonlinear asymmetric devices typically incorporate nonlinear materials into structures with direction-dependent electric-field distributions. When light of sufficient intensity is incident upon such a structure from one direction, the electric field within the nonlinear material results in a change in the transmission. Light of the same intensity incident from the opposite direction generates a different field distribution that does not trigger the nonlinearity, resulting in the asymmetry. Because optical nonlinearities are small for most materials [16][17], high intensities are typically needed to achieve substantially asymmetric transmission. If the input intensity is not sufficiently high, optical cavities with large quality factors can be used to enhance the field, but this results in a narrow working bandwidth [18] and may require complex geometries [3][18][19].

Recently, vanadium dioxide ($VO_2$) has been explored as a tunable material for optics applications due to the large change in complex refractive index across its insulator-metal transition (IMT) [20][21][22]. This change in the refractive index is orders of magnitude larger than that of the Kerr effect and similar nonlinearities, particularly in the near infrared and longer wavelengths (*e.g.*, at $\lambda$ = 1.55 µm, $\Delta n_{VO2}$ ~ 1.5 [21][23] vs. $\Delta n$ ~ 0.0001 for silicon given an intensity of ~3× $10^6$ kW/cm$^2$ [16]), making $VO_2$ a promising nonlinear material for limiting optical diodes and other nonlinear isolators.

In this letter, we design, model, fabricate, and characterize a thin-film near-infrared limiting optical diode based on the photothermally driven IMT in $VO_2$. Our strategy is to utilize asymmetric absorption within a thin film of $VO_2$ to selectively trigger the IMT, enabling asymmetric transmission.

**Experimental results**

Our proof-of-concept device is a simple thin-film stack comprising a transparent sapphire substrate, a thin $VO_2$ layer, and a semi-transparent gold film (Fig. 1(a)). When light is incident from the gold side ("forward incidence"), a significant amount of power is reflected before it reaches the $VO_2$. When light is incident from the sapphire side ("backward incidence"), there is substantially more absorption of the light by the $VO_2$ compared to the forward-incidence case. Thus, there exists a range of incident intensities for which the photothermal heating is sufficient to trigger the IMT, minimizing optical transmission, for backward incidence only. Note that the gold thickness must be judiciously chosen, because a thicker film increases the absorption asymmetry but also decreases the total transmission in the transparent state of the device.

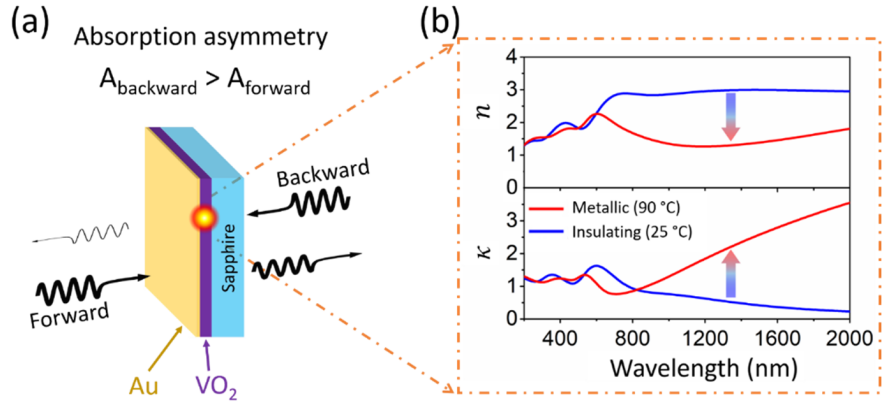

**Figure 1**. **(a)** Schematic of our planar limiting optical diode, in which the onset of the metallic phase of $VO_2$ is triggered by intense "backward" illumination, reducing transmission through the device. The $VO_2$ maintains its insulating phase when illuminated in the forward direction at the same illumination intensity, resulting in high transmission. **(b)** Real and imaginary refractive indices of our sputtered $VO_2$ film (extracted using spectroscopic ellipsometry) in its insulating (25 °C) and metallic (90 °C) phases. At our experimental wavelength $\lambda$ = 1.32 μm, the complex refractive index changes from $3.2 + 0.6i$ to $1.7 + 2.4i$ across the phase transition.

The temperature-dependent infrared refractive indices of the $VO_2$ film were obtained using variable-angle spectroscopic ellipsometry (J. A. Woollam Co. IR-VASE Mark II) with a temperature-controlled sample stage (Fig. 1(b)). Then, using the extracted optical constants, $VO_2$ and gold film thicknesses were optimized using transfer-matrix calculations, finding a reasonable tradeoff for two figures of merit: absorption asymmetry and transmission in the low-power state. Our calculations (see Supplementary Section 1 for details) show that using a 100-nm-thick $VO_2$ layer sandwiched between a 10 nm evaporated gold layer and a sapphire substrate, an absorption asymmetry ratio of ~6.7 can be achieved when the $VO_2$ is in its insulating state (absorption of ~0.1 for forward illumination, versus ~0.67 for backward illumination). This structure features a reciprocal transmission of ~0.22 in the low-power state, and a non-reciprocal high-power-state transmission of ~0.22 for forward incidence and ~0.04 for backward incidence. At even higher powers, the transmission again becomes reciprocal (~0.04 from both sides, Supplementary Section 4).

We fabricated the device using two deposition steps. First, $VO_2$ was deposited onto a two-side polished *c*-plane sapphire wafer via magnetron sputtering from a $V_2O_5$ target, with radio-frequency power of 100 W. During deposition, the chamber pressure was maintained at 5 mTorr with an $Ar/O_2$ gas mixture at a flow rate of 49.9/0.1 sccm. The substrate was heated to 650 °C for the formation of the $VO_2$ phase [24]. After deposition, scanning electron microscopy (SEM) and atomic force microscopy (AFM) characterization confirmed the film was continuous, with an average surface roughness $R_a$ ~ 4.1 nm (Supplementary Section 2). Then, 10 nm of gold was evaporated on top of the $VO_2$ using an electron-beam evaporator.

We tested our device using a custom-built transmission setup (Fig. 2(a)), using a continuous-wave laser operating at λ = 1.32 μm with a maximum output power of ~40 mW (Thorlabs TLK-L1300R). A half-wave plate and polarizer were used as a variable neutral-density (ND) filter, with at least three orders of magnitude attenuation available. The incident beam was focused (NA ~0.1) onto our limiting optical diode, resulting in a maximum incident intensity of ~9 kW/cm$^2$. The diode was mounted on a hot stage with a through-hole aperture, used to thermally bias the sample. This thermal biasing was necessary because the maximum focused laser intensity was too low to photothermally heat the $VO_2$ layer from room temperature to the IMT temperature. Light passing through the device was focused onto a germanium detector (Thorlabs S132C) to measure the transmission, which was collected for temperatures from below to above the IMT temperature. Efforts were made to avoid the well-known hysteresis in $VO_2$ [20][25] from influencing our measurements, by heating the sample from room temperature to a set bias temperature for each data point collected. The transmitted power was recorded once the bias temperature was reached.

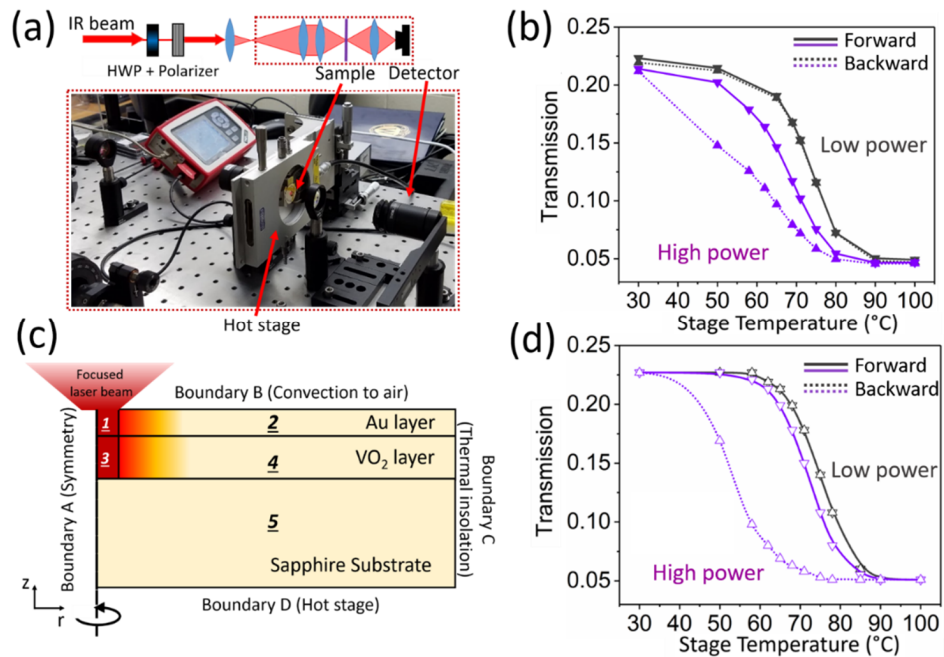

**Figure 2. (a)** Transmission measurement using a continuous-wave laser (λ = 1.32 μm) and a hot stage for thermal biasing. **(b)** Measured transmission for "forward" and "backward" illumination as a function of temperature for low (0.5 kW/cm$^2$) and high (9 kW/cm$^2$) incident intensity. **(c)** Geometry for photothermal simulations. The axisymmetric model consists of five domains for the three layers, of which domains 1 and 3 experience laser absorption. **(d)** Simulated transmission approximately corresponding to the experiments in (b) using a coupled optical and thermal model.

The temperature-dependent transmission from both incidence directions is shown in Fig. 2(b), given high (~9 kW/cm$^2$) and low (~0.5 kW/cm$^2$) laser intensity. For low intensity, no asymmetric transmission was observed at any bias temperature. This indicates that from either direction, the absorbed power is insufficient to trigger the IMT. However, in the case of high incident intensity, the transmission shows a clear directional dependence (purple curves in Fig. 2(b)). For example, when the hot-stage bias temperature ($T_{stage}$) is set to 50 °C (below the IMT temperature), the sample transmission is ~0.21 for forward incidence, but ~0.15 for backward incidence. The decrease in transmission for backward incidence corresponds to a VO$_2$ temperature of approximately 71 °C, while for forward incidence it corresponds to a much lower temperature of ~54 ºC, due to the smaller amount of photothermal heating. Note that for each bias temperature in Fig. 2(b), the VO$_2$ only partially transitioned to the metallic phase due to the relatively wide phase-transition temperature range of our VO$_2$ films (see Supplementary Section 4). The highest transmission asymmetry obtained experimentally is ~2 at $T_{stage}$ = 69 °C.

**Simulation and discussion**

To further understand the working mechanism of our diode, the photothermally driven refractive-index change in VO$_2$ must be carefully considered. As shown in Fig. 1(b), VO$_2$ becomes more absorbing when it transitions to its metallic phase. Thus, as the incident light is absorbed, the film is gradually heated, increasing $\kappa$, enabling it to absorb more power and heat up further. This positive feedback mechanism drives the VO$_2$ film further into the metallic phase until equilibrium is reached.

We constructed a model that couples thermal conduction (using COMSOL Multiphysics) and optical absorption (using the transfer-matrix method) to simulate this behavior. We briefly outline our method here, with more details in Supplementary Section 3. The thermal-conduction model consists of an axisymmetric geometry of five domains centered on the center point of the laser beam, which provided an axisymmetric boundary (Fig. 2(c)). The temperature evolution follows the general steady-state heat equation [26], and we assumed the thermal conductivity and specific heat capacity of VO$_2$ to be 6 W/(m·K) [27] and 690 J/(K·kg) [28], respectively. Note that these values correspond to VO$_2$ in its insulating phase, but their change due to the IMT is not dramatic (less than 20% across the IMT) [27][28], so for simplicity we set them to be constant. Thermal properties of Au and sapphire were taken from Refs. [29] and [30]. We assumed that the heat generation within the device is entirely due to laser power absorbed in the Au and VO$_2$ layers (domains 1 and 3 in Fig. 2(c)). The laser beam used in our model had a Gaussian distribution with a full-width at half maximum (FWHM) of 24 µm, matching our experiments. The absorption within the VO$_2$ layer at each spatial position $(r, z)$ depends on the incident intensity as well as the local temperature, via the temperature-dependent refractive index. The full temperature-dependent complex refractive index of VO$_2$ for temperatures between 30 °C and 90 °C was calculated using the insulating- and metallic-phase values shown in Fig. 1(b), and applying an effective-medium approximation [31]. The temperature-dependent absorption within the VO$_2$ and Au layers was calculated via the transfer-matrix method and then plugged into our heat-transfer calculation as input parameters. The simulation boundary conditions included $T(z = 0) = T_{stage}$ ("D" in Fig. 2(c)), and a convective boundary to ambient air at the top and side surfaces ("B" and "C" in Fig. 2(c)).

Our iterative model works as follows to calculate the transmission given a bias temperature and a particular incident intensity for a given direction of incidence. The complex refractive index of $VO_2$ for $T = T_{stage}$ is used to calculate the optical absorption and the resulting temperature distribution within the Au and $VO_2$ layers. This temperature distribution is used to update a refractive-index profile, which is then fed into the next simulation step. This process is iterated until the temperature distribution converged to its equilibrium value.

Our model captured the trend of our experimental results (Fig. 2(d) vs 2(b)), with some discrepancies. To reasonably match the experimental data, we used an incidence intensity of ~6 kW/cm$^2$, compared to the experimental intensity of ~9 kW/cm$^2$. Also, the shift of the high-power backward transmission to lower temperatures was more pronounced in the simulation data, resulting in a maximal asymmetric transmission ratio of 4.1 at $T_{stage}$ = 58 °C; by comparison, the maximal experimental asymmetry was measured to be ~1.95 at $T_{stage}$ = 69 °C. These discrepancies may be a result of some simplified assumptions we made in our model. For example, we ignored the change in the thermal conductivity and heat capacity of $VO_2$ across the IMT, and we did not consider the effect of surface roughness at all interfaces. It is also possible that our focusing was less than ideal, resulting in a slightly lower incidence intensity in the experiment.

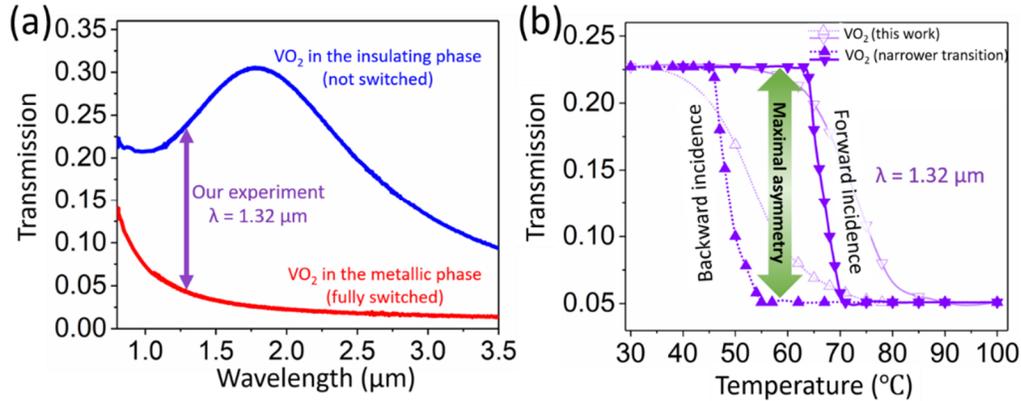

**Figure 3. (a)** Transmission spectra of our limiting optical diode in the unswitched (blue) and switched (red) state, measured at room temperature and thermally biased to 90 °C, respectively. **(b)** Simulation results corresponding to Fig. 2(d), and also including $VO_2$ with a narrower phase transition (*e.g.*, the films in Ref. [32]). For detailed description of optical properties of both types of $VO_2$, see Supplementary Section 4.

Because $VO_2$ can experience a large change in $n$ and $\kappa$ given a relatively small change in temperature (and hence a small amount of input power), a limiting optical diode that uses $VO_2$ as an active material does not necessarily require field-enhancement structures like the Fabry-Perot-type cavities used for diodes based on conventional nonlinearities [3][18]. As a result, a $VO_2$-based limiting optical diode can be very thin, and can be designed to operate over a broad wavelength range. For example, our planar diode has no sharp features in its transmission spectrum over the 1 – 3.5 *μ*m range (Fig. 3(a)).

The contrast and threshold power of our device may be improved by using a phase-transition material with a sharper transition. In our device, the transmission asymmetry is limited by the relatively broad phase transition in our multi-domain $VO_2$ film grown on a *c*-plane sapphire [33][34]. The broad temperature range over which the IMT occurs means that a very large absorption asymmetry is necessary to fully switch the $VO_2$ film to the metallic phase for one incidence direction, but leave the film in its insulating phase for the other direction; in fact, we did not accomplish this in our experiments (Fig. 2(b)). Our simulations predict that by using $VO_2$ films with narrower transitions [32][35][36][37], maximal asymmetry can be reached in our simple thin-film geometry for the same intensity as in our experiment (Fig. 3(b); for details see Supplementary Section 4).

It is also important to consider the effect of the hysteresis intrinsic to $VO_2$ [25] on the operation of $VO_2$-based limiting optical diodes. For most applications, the diode should automatically revert to its transparent state once the light driving the transition is turned off. Therefore, the device should not be thermally biased at any temperature within the hysteresis loop. This is a more restrictive but still reachable condition for maximal transmission asymmetry (see Supplementary Section 5). Finally, it is worth mentioning that techniques, such as defect engineering [31] or doping [38][39] of the $VO_2$ film enable the phase transition to be shifted closer to room temperature, which allows for a controllable power threshold and a completely passive device that does not require thermal biasing.

**Summary**

We have demonstrated a limiting optical diode based on asymmetric optical absorption in a vanadium dioxide ($VO_2$) film embedded in a simple thin-film assembly. The large change in complex refractive index across the insulator-to-metal transition (IMT) in $VO_2$ makes it possible to design limiting optical diodes and other nonlinear isolators without field-enhancing architectures, minimizing thickness and maximizing operating bandwidth. As a result, our simple device (which reached an experimental transmission asymmetry of ~2) is broadband and approximately ten times thinner than the operating wavelength.

**Acknowledgements**

We acknowledge support from the Office of Naval Research (Grant No. N00014-16-1-2556, and N00014-16-1-2398) and National Science Foundation (Grant No. DMR-1610345). We thank Zongfu Yu, Jura Rensberg, and Carsten Ronning for helpful discussions. Some of the fabrication and experiments were performed at the Materials Science Center (MSC) and the Wisconsin Center for Applied Microelectronics (WCAM), both shared facilities managed by College of Engineering at the University of Wisconsin-Madison.**References**

[1]  D. Jalas *et al.*, "What is-and what is not-an optical isolator," *Nature Photonics*, vol. 7, no. 8, pp. 579–582, 2013.

[2]  L. Bi *et al.*, "On-chip optical isolation in monolithically integrated non-reciprocal optical resonators," *Nature Photonics*, vol. 5, no. 12, pp. 758–762, Dec. 2011.

[3]  M. D. Tocci, M. J. Bloemer, M. Scalora, J. P. Dowling, and C. M. Bowden, "Thin-film nonlinear optical diode," *Applied Physics Letters*, vol. 66, no. 18, pp. 2324–2326, 1995.

[4]  A. M. Mahmoud, A. R. Davoyan, and N. Engheta, "All-passive nonreciprocal metastructure," *Nature Communications*, vol. 6, p. 8359, Sep. 2015.

[5]  M. Kimura, K. Onodera, and T. Masumoto, "980 nm compact optical isolators using $Cd_{(1-x-y)}Mn_xHg_yTe$ single crystals for high power pumping laser diodes," *Electronics Letters*, vol. 30, no. 23, pp. 1954–1955, Nov. 1994.

[6]  J. Z. Yuanxin Shou, Qingdon Guo, "High power optical isolator," US7715664 B1, 29-Oct-2008.

[7]  L. J. Aplet and J. W. Carson, "A Faraday Effect Optical Isolator," *Applied Optics*, vol. 3, no. 4, p. 544, 1964.

[8]  H. Dötsch *et al.*, "Applications of magneto-optical waveguides in integrated optics: review," *Journal of the Optical Society of America B*, vol. 22, no. 1, p. 240, 2005.

[9]  Y. Shoji, T. Mizumoto, H. Yokoi, I.-W. Hsieh, and R. M. Osgood, "Magneto-optical isolator with silicon waveguides fabricated by direct bonding," *Applied Physics Letters*, vol. 92, no. 7, p. 71117, Feb. 2008.

[10] P. Weinberger, "John Kerr and his effects found in 1877 and 1878," *Philosophical Magazine Letters*, vol. 88, no. 12, pp. 897–907, 2008.

[11] N. J. White *et al.*, "Nonreciprocal Light Propagation in a Silicon Photonic Circuit," *Science*, vol. 333, no. August, pp. 729–733, 2011.

[12] M. Soljačić, C. Luo, J. D. Joannopoulos, and S. Fan, "Nonlinear photonic crystal microdevices for optical integration," *Optics Letters*, vol. 28, no. 8, p. 637, Apr. 2003.

[13] Y. Shi, Z. Yu, and S. Fan, "Limitations of nonlinear optical isolators due to dynamic reciprocity," *Nature Photonics*, vol. 9, no. 6, pp. 388–392, 2015.

[14] M. Cronin☐Golomb and A. Yariv, "Optical limiters using photorefractive nonlinearities," *Journal of Applied Physics*, vol. 57, no. 11, pp. 4906–4910, Jun. 1985.

[15] B. L. Justus, A. L. Huston, and A. J. Campillo, "Broadband thermal optical limiter," *Applied Physics Letters*, vol. 63, no. 11, pp. 1483–1485, Sep. 1993.

[16] J. Leuthold, C. Koos, and W. Freude, "Nonlinear silicon photonics," *Nature Photonics*, vol. 4, no. 8, pp. 535–544, 2010.8

# Supplementary information:

# Limiting optical diodes enabled by the phase transition of vanadium dioxide

Chenghao Wan[1,2], Erik Horak[3], Jonathan King[1], Jad Salman[1], Zhen Zhang[4], You Zhou[5], Patrick Roney[1], Bradley Gundlach[1], Shriram Ramanathan[4], Randall Goldsmith[3], Mikhail A. Kats[1,2]

[1] Department of Electrical and Computer Engineering, University of Wisconsin-Madison, Madison, Wisconsin 53706, USA
[2] Department of Materials Science & Engineering, University of Wisconsin-Madison, Madison, Wisconsin 53706, USA
[3] Department of Chemistry, University of Wisconsin-Madison, Madison, Wisconsin 53706, USA
[4] School of Materials Engineering, Purdue University, West Lafayette, IN 47907, USA
[5] Harvard John A. Paulson School of Engineering and Applied Sciences, Harvard University, Cambridge, MA 02138, USA

**Section 1. Optimization of forward transmission and absorption asymmetry**

To balance between the two figures of merit (maximized absorption asymmetry and sufficient transmission in the low-power state), we swept the thicknesses of Au and $VO_2$ and calculated the transmission and asymmetry ratio of absorption using the transfer-matrix method, as shown in Fig. S-1. The same *c*-plane sapphire substrate was used in all calculations. Thicker $VO_2$ and Au generally results in larger absorption asymmetry, but also lower transmission (Fig. S-1(a)). We chose to fabricate a sample with 100 nm $VO_2$ and 10 nm Au, resulting in transmission of ~0.22 at low power, and a six-fold absorption asymmetry (absorption = 0.11 for forward propagation, and 0.67 for backward propagation).

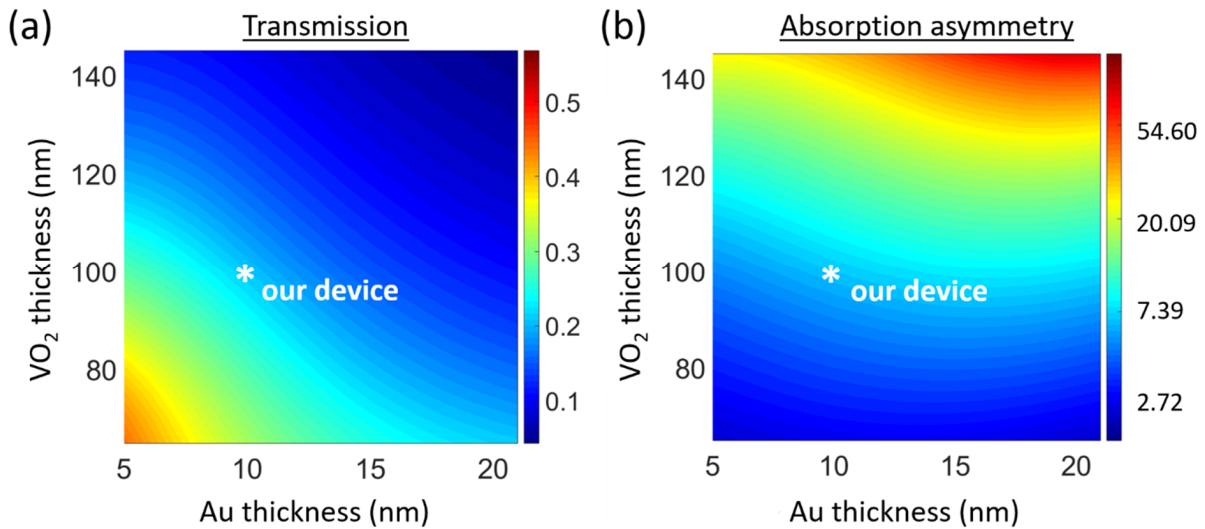

**Figure S-1. (a)** Transmission and **(b)** absorption-asymmetry ratio ($A_{backward}/A_{forward}$) of our thin-film device with the thickness of $VO_2$ varying from 60 to 150 nm, and the thickness of Au varying from 5 to 21 nm. The refractive index of $VO_2$ used in these calculations corresponds to its insulating phase.

**Section 2. SEM and AFM characterization of our sputtered $VO_2$ film on c-plane sapphire**

To examine the surface uniformity of the $VO_2$, we characterized the post-growth film using an SEM (Zeiss LEO 1530) and an AFM (Veeco MultiMode SPM) in tapping mode (Fig. S-2). Based on five AFM images, each of which is 10 µm by 10 µm (*e.g.*, Fig. S-2(b)), we calculated the surface roughness to be Ra = 4.11 nm.

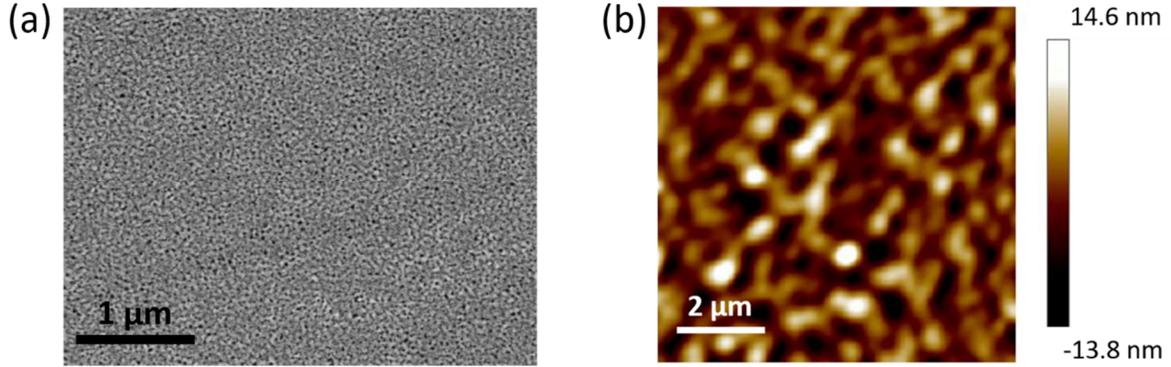

**Figure S-2. (a)** SEM and **(b)** AFM images of post-growth 100 nm $VO_2$ on a c-plane sapphire substrate. A surface roughness $R_a$ = 4.11 nm was calculated based on five AFM images, each of which is 10 µm by 10 µm.

**Section 3. Details of our model**

Spectroscopic ellipsometry (SE) was used to extract the complex refractive indices of $VO_2$ in the insulating (30 °C) and metallic phase (90 °C), for free-space wavelengths from 200 nm to 2000 nm. In our fitting model, optical constants of $VO_2$ are described by a general oscillator (GNO) function [S1], which includes five Gaussian oscillators [S2] for the insulating phase, and an additional Drude term [S2] for the metallic phase. The results are shown in Fig. 1(b) in the main text. An effective medium approximation (Looyenga mixing rule; Eq. (S1)) [S3][S4] was then used to estimate the refractive indices of $VO_2$ across its phase transition, as shown in Fig. S-3(a).

$$\tilde{\varepsilon}_{eff}^{\ m} = (1-f)\tilde{\varepsilon}_{insulating}^{\ m} + f\tilde{\varepsilon}_{metallic}^{\ m} \quad (S1)$$

$$\tilde{n} = n + i\kappa = \sqrt{\tilde{\varepsilon}} \quad (S2),$$

where $\tilde{\varepsilon}$ is the complex dielectric function of $VO_2$, $f$ is the metallic fraction, and $m$ is an exponent that describes the anisotropy of the material. $f$ and $m$ were both estimated using the same method as in Ref. [S4]. The effective optical refractive index was then calculated using Eq. (S2). For our experimental wavelength ($\lambda$ = 1.32 µm), there is a significant change in the complex refractive index of $VO_2$, from $\tilde{n}$ =

3.2 + 0.6$i$ to $\tilde{n}$ = 1.7 + 2.4$i$ across its phase transition. The temperature-dependent refractive index of VO$_2$ (Fig. S-3(a)) enabled us to calculate the absorption within our device for both forward and backward incidence as a function of temperature (Fig. S-3(b)). These absorption curves were used as input parameters for our COMSOL heat-transfer model.

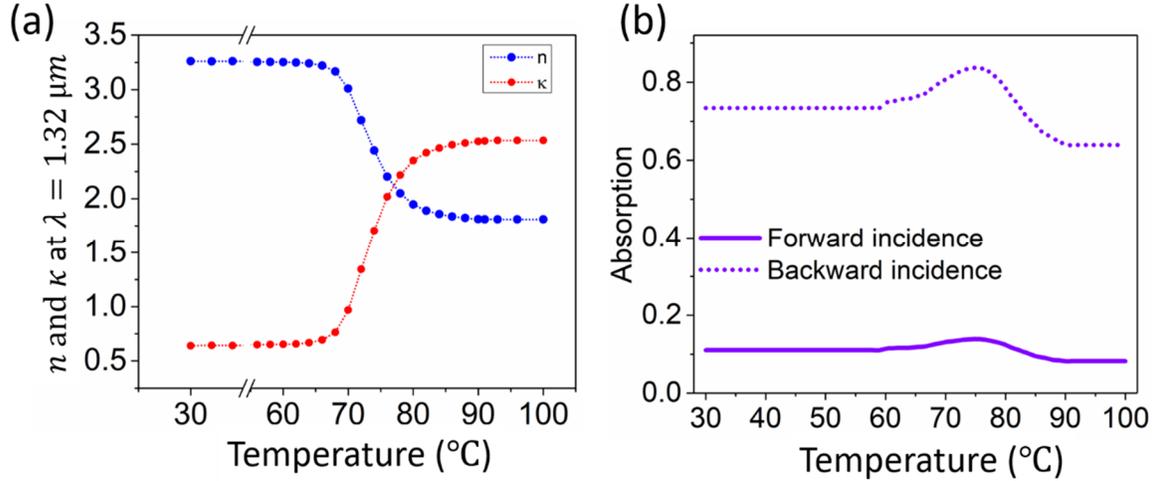

**Figure S-3. (a)** Calculated refractive index of VO$_2$ (at λ = 1.32 μm) from T = 30 to 100 °C, using the Looyenga mixing rule. **(b)** Calculated combined absorption in the VO$_2$ and Au layers (region 1 and 3 in Fig. 2(c)) as a function of the bias temperature.

In our heat-transfer model, the temperature evolution of the structure follows the general steady-state heat equation (Eq. (S3)) [S5], by which the generated heat ($Q$) is proportional to the local heat-flux density ($\bm{q}$):

$$-\nabla \cdot \bm{q} + Q(r,z,T) = 0 \qquad (S3),$$

where

$$\bm{q} = -k\nabla T(r,z) \qquad (S4)$$

$$Q(r,z,T) = A(T(r,z)) \qquad (S5)$$

Eq. (S4) is Fourier's law, which describes that the heat-flux density is proportional to the gradient of temperature ($T$), where $k$ is the thermal conductivity. Eq. (S5) assumes that the heat-generation term ($Q$) is based on the amount of laser power absorbed in each layer ($A$). The absorption within the VO$_2$ layer is a function of the varying local temperature at each spatial position $(r,z)$, and is given in Fig. S-3(b).

## Section 4. Proof-of-concept design using $VO_2$ with a narrower phase transition

The relatively broad temperature range of the transition of the $VO_2$ films used in our experiments is a potential limitation of our device. For the maximum laser intensity we could experimentally achieve, there was no bias temperature for which the diode is maximally asymmetric; *i.e.,* the transmission is maximal for light incident from one side, and minimal for light incident from the other side. To realize maximal transmission asymmetry at the intensities in our experiment, one could use $VO_2$ with a narrower phase transition, *e.g.*, using other substrates or engineered strain conditions for oriented growth [S6][S7]. For comparison, the solid-point curve in Fig. S-4(a) is the calculated temperature-dependent transmission, generated based on Ref. [S6], assuming that the material has the same range of optical properties as that used in our work, but with a narrower phase-transition range (~6 ºC, from 65 ºC to 71 ºC). The open-point curve in Fig. S-4(a) is the temperature-dependent transmission of our fabricated device. We used our model to investigate the diode performance of both designs. As shown in Fig. S-4(b), our simulation indicates that the design based on narrower-transition $VO_2$ can reach maximal transmission asymmetry when it is thermally biased between 55 and 63 ºC, for the incident intensities described in the main text.

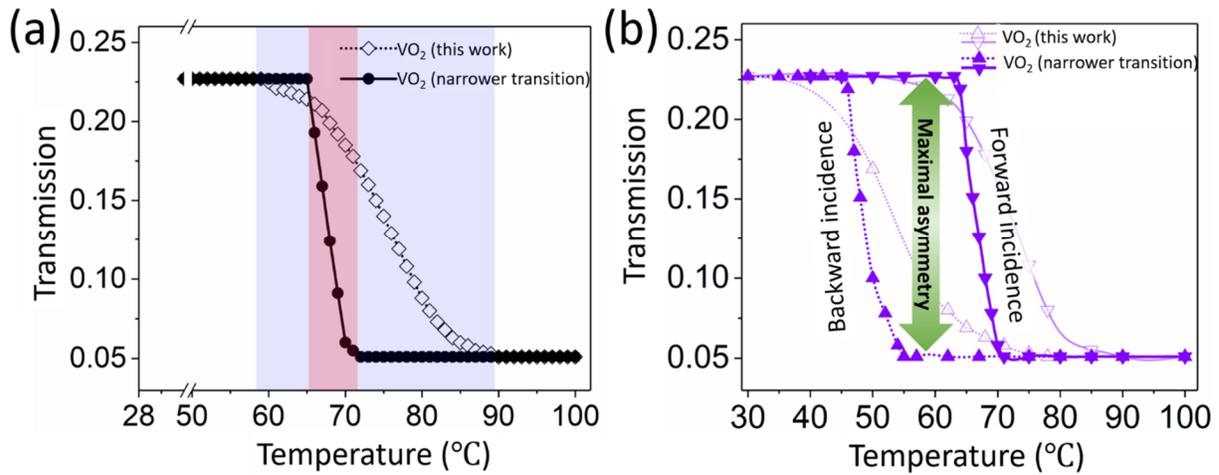

**Figure S-4. (a)** Temperature-dependent (during heating) transmission at λ = 1.32 μm of our fabricated limiting optical diode, and calculated transmission of the same structure but using the $VO_2$ from Ref. [S6], assuming it has the same optical properties as our sputtered $VO_2$ but with a narrower phase transition. **(b)** Limiting diode performance simulated using our model, given the experimental conditions described in the main text. Duplicate of Fig. 3(b) in the main text.

Furthermore, our simulations show that our fabricated device—without any modification—will reach maximal asymmetry if driven at higher intensities than those realized in our experiment (Fig. S-5(a)). We calculated the intensity-dependent transmission-asymmetry ratio for a thermal-bias temperature at 50 ºC, assuming both broader- and narrower-transition $VO_2$ in our limiting diodes. We also investigated the performance at different bias temperatures (Fig. S-5(b)). Note that in Fig. S-5(b), we only show the results for the design based on narrower-transition VO$_2$. As expected, higher incident intensities are needed to trigger the device given lower bias temperatures. If the incident intensity is high enough, our design can work passively without any thermal bias.

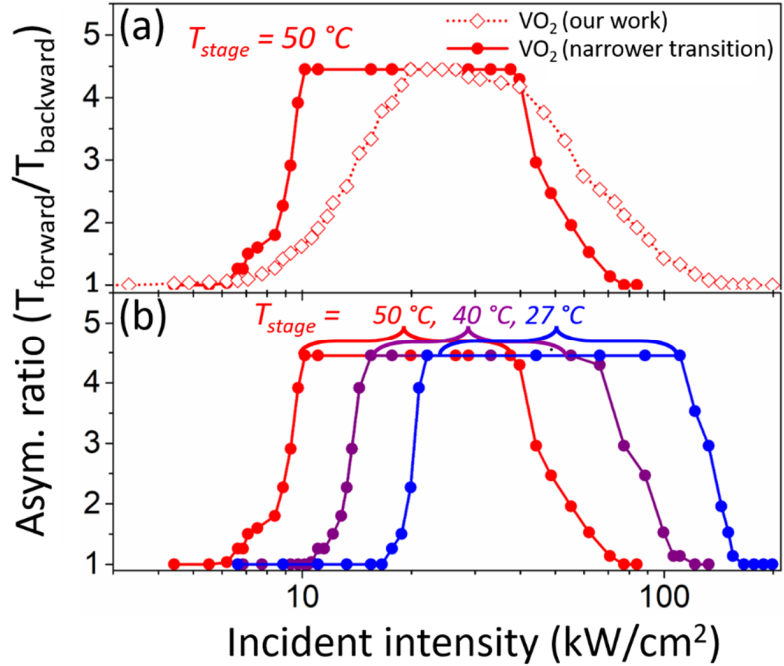

**Figure S-5.** Simulated transmission asymmetry at varying incident intensities. **(a)** At $T_{stage}$ = 50 °C, comparison of the asymmetry ratio between identical devices based on narrower-transition VO$_2$ and the VO$_2$ used in our experiments. VO$_2$ with a narrower transition range enables a broader intensity range at which the asymmetry ratio is maximized. **(b)** Intensity-dependent asymmetry ratio of the design based on narrower-transition VO$_2$ at $T_{stage}$ = 27 °C, 40 °C, and 50 °C.

**Section 5. Hysteresis discussion**

In practice, one needs to consider the hysteresis effect in VO$_2$, as shown in Fig. S-6(a) for narrower-transition VO$_2$ from Ref. [S6], and Fig. S-6(c) for the VO$_2$ used in our experiments. To ensure that the diode is automatically reset when the illumination is turned off, one needs to set the bias temperature below the hysteresis region for the unilluminated diode. As identified using red shading in Fig. S-6(a), suitable thermal-bias temperatures range from room temperature to 56 °C for the design based on narrower-transition VO$_2$. Given an input intensity of 6 kW/cm$^2$ (the same simulation condition as in Fig. S-4(b)), one can achieve maximal asymmetry at bias temperatures from 55 °C to 64 °C (identified using green shading in Fig. S-6(b)). The region in common, where the green and red overlap, is the optimal operating range of the device. (Fig. S-6(a, b)). For our fabricated device (Fig. S-6(c), (d)) given the same incident intensity, there are also suitable bias temperatures (identified using red shading in Fig. S-6(c)) at which asymmetric transmission can be realized, though the asymmetry ratio is not maximal. As our model predicts (Section 4), a higher incident intensity is necessary for our fabricated device to realize maximal asymmetry.

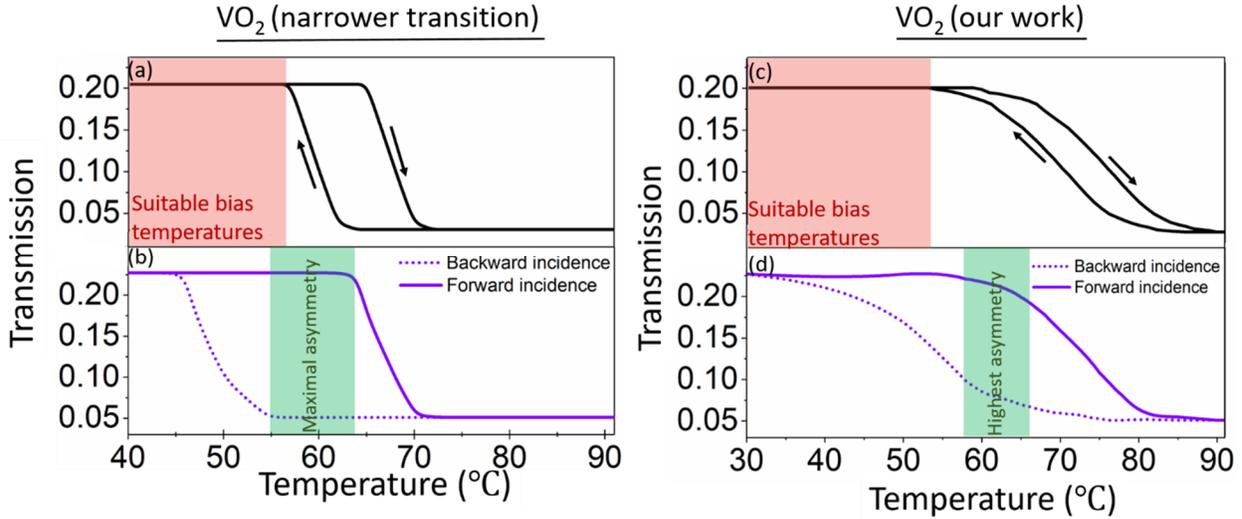

**Figure S-6. (a)** Hysteresis loop in the transmission of the device based on narrower-transition $VO_2$. The plot was generated using the same model as in Section 4, assuming no illumination. Temperatures within the red-shaded region are suitable to fully avoid device hysteresis. **(b)** Simulation results for the same diode design based on narrower-transition $VO_2$ for the high-power-incidence case (6 kW/cm$^2$, the same scenario as in Fig. S-4(b)). Temperatures within the green-shaded region yield maximal transmission asymmetry. **(c)** and **(d)** are the corresponding calculations for the design based on the $VO_2$ used in our experiments.